# Angular-Momentum Selectivity and Asymmetry in Highly Confined Wave Propagation Along Sheath-Helical Metasurface Tubes

Yarden Mazor[1], and Andrea Alù[1,2,3,4*]

[1]Department of Electrical and Computer Engineering, The University of Texas at Austin, Austin, TX 78712, USA

[2]Photonics Initiative, Advanced Science Research Center, City University of New York, New York, NY 10031, USA

[3]Physics Program, Graduate Center, City University of New York, New York, NY 10026, USA

[4]Department of Electrical Engineering, City College of New York, New York, NY 10031, USA

*Highly confined surface waves present unique opportunities to enhance light interactions with localized emitters or molecules. Hyperbolic dispersion in metasurfaces allows us to tailor and manipulate surface waves, enhancing the local density of states over broad bandwidths. So far, propagation on this platform was mainly studied in planar geometries, which facilitates the analysis but somehow limits the realm of possibilities. Here we show that "wrapping" hyperbolic metasurfaces into tubes may greatly enrich the wave propagation dynamics along their axis. This system shows strong interaction with fields and sources carrying optical angular momentum, pronounced field asymmetries, and opens pathways to valley-specific excitation and routing. In addition, we demonstrate that various parameter regimes enable strong spin/helicity – momentum locking.*

1. **Introduction**

Propagation of surface waves along planar structures has been a vastly studied subject over time. In the past decade, this subject has seen increased interest in the context of plasmonics and the emergence of metasurfaces [1] – thin sheets of matter with a carefully engineered response to electromagnetic fields offering new opportunities to control and manipulate the fields. Among the numerous applications metasurfaces may have, more relevant to the present work is the research unveiling new phenomena in surface wave propagation. To model the metasurface response, the surface impedance concept is used in many cases [2]. Propagation of surface waves on impedance metasurfaces has been extensively studied [3–5], presenting the flexibility offered by tailoring the surface impedance components to control and steer the surface wave fields. By considering additional degrees of freedom to the metasurface response, such as anisotropy and bianisotropy [6] we can further tailor the overall reflection/transmission [7,8] and propagation characteristics [4,9]. These additional parameters can lead to exotic responses, such as hyperbolic metasurfaces [10–13] that support propagation of surface waves with a hyperbolic dispersion, allowing highly confined guiding and Purcell enhancement over broad bandwidths. Magnetic field biasing also allows nonreciprocal sector-way guiding [14]. Combined with modern nanofabrication techniques, graphene flexibility and tunability opens a realistic route towards implementing these phenomena in practical devices [11].

Surface waves guided along cylindrical boundaries have also been researched, following the pioneering work [15] that introduced wave propagation along the boundary of a metallic cylinder with finite conductivity. This study was extended in [16], both elaborating on the case of a smooth boundary, and discussing a corrugated one. In [17] the plasma response of the metal was also considered as a model of the material response. Azimuthal leaky surface waves, which propagate in the $\hat{\varphi}$ cylindrical direction rather than in the axial one, were also studied in [18,19] quantifying their radiation properties. A different classical geometry, from which our work draws some of its roots, is based on perfectly conducting sheath helices [20,21] - allowing current only at a specific angle with respect to the cylinder axis. These structures support axial propagation of higher order, highly confined circular modes carrying optical angular momentum (OAM) of the form $e^{in\varphi}$, where a mode with specific $n$ has asymmetric guiding properties. Upon the emergence of carbon nanotubes, a model of a cylindrical impedance surface was employed for several configurations [22,23] to study their electrodynamics and wave guiding properties, a model which we later use here. Cylindrical and spherical sheath metasurfaces have also been considered as candidates for cloaking [24–26] and engineering nanoparticle resonant response [27]. Cylindrical surface wave propagation was also considered in [28] for a dispersive plasma cylinder model, were it was shown that all modes converge towards specific frequencies associated with the plasma and magnetic resonance frequencies. Scattering from such cylinders was also studied in [28,29]. With recent advances in manufacturing techniques, we envision more complex cylindrical surfaces made of carefully designed and arranged inclusions. In the context of scattering, Ref. [30] lays the foundations for the analysis of scattering from this type of cylindrical bianisotropic metasurface, where a method to extract the required metasurface design for a prescribed response is discussed.

In this work, we study planar hyperbolic metasurfaces rolled into a cylindrical tube in various configurations, yielding a cylindrical sheath with opposite current responses in two perpendicular directions. The modes propagating along the cylinder axis present extreme dispersion asymmetry. We show that by incorporating a magnetic response, the asymmetries can be enhanced, and asymmetric field distributions localized inside and outside the cylinder can be obtained. We examine propagating waves in the context of the electromagnetic helicity, and explore different helicity regimes for different surface parameters. These cylindrical metasurfaces reveal opportunities for highly asymmetric interaction with OAM / spin carrying waves such as unidirectional excitation of modes.

## 2. Geometry of the problem

The geometry under consideration is shown in figure (1). To intuitively describe the surface response, we may relate it to a planar impedance surface folded into a cylinder, where the cylinder axis makes an angle $\theta$ with the principal axes (red and blue lines) of the impedance tensor. We assume the surrounding medium to be free-space. The $\hat{\mathbf{z}}$ components of the EM fields guided along the cylinder can be written as [31]

$$E_z = \begin{Bmatrix} A_n^i I_n(\tau r) \\ A_n^o K_n(\tau r) \end{Bmatrix} e^{-j\beta z} e^{jn\varphi} \quad , \quad H_z = \begin{Bmatrix} B_n^i I_n(\tau r) \\ B_n^o K_n(\tau r) \end{Bmatrix} e^{-j\beta z} e^{jn\varphi}, \tag{1}$$

where $A_n^i$ and $B_n^i$ are the amplitudes of transverse-magnetic (TM) and transverse-electric (TE) wave inside the cylinder, and $A_n^o$, $B_n^o$ outside the cylinder. $I_n$ and $K_n$ are the modified Bessel functions of the 1st and 2nd kind. For surface waves to be guided, there needs to be a real $\beta$ solution satisfying $\beta^2 = k_0^2 + \tau^2 > k_0^2$, with $k_0 = \omega\sqrt{\varepsilon_0\mu_0} = \omega/c$ the free-space wavenumber. In the dispersion curve, such a mode would appear below the light lines which correspond to propagation with wavenumber $k_0$. The other field components are calculated from equation (1) using Maxwell's equations (see Appendix 1). The properties of the impedance surface are taken into account through the boundary conditions

$$\hat{\mathbf{r}} \times (\mathbf{H}^o - \mathbf{H}^i)_{r=a} = \mathbf{J}_s = \frac{1}{2} \underline{\underline{\mathbf{Y}}}_s \left( \mathbf{E}_{\tan}^i + \mathbf{E}_{\tan}^o \right) + \frac{1}{2} \underline{\underline{\mathbf{a}}} \left( \mathbf{H}_{\tan}^i + \mathbf{H}_{\tan}^o \right), \tag{2}$$

$$-\hat{\mathbf{r}} \times \left( \mathbf{E}^o - \mathbf{E}^i \right) = \mathbf{J}_{ms} = \frac{1}{2} \underline{\underline{\mathbf{Z}}}_{ms} \left( \mathbf{H}^i_{\tan} + \mathbf{H}^o_{\tan} \right) + \frac{1}{2} \underline{\underline{\mathbf{b}}} \left( \mathbf{E}^i_{\tan} + \mathbf{E}^o_{\tan} \right). \qquad (3)$$

Here, $\mathbf{J}_s, \mathbf{J}_{ms}$ are the tangential electric and magnetic currents on the surface of the cylinder, $\underline{\underline{\mathbf{Y}}}_s$ is the electric surface admittance, $\underline{\underline{\mathbf{Z}}}_{ms}$ the magnetic surface impedance, $\underline{\underline{\mathbf{a}}}, \underline{\underline{\mathbf{b}}}$ are the bianisotropic electric-magnetic coupling tensors, and the suffix *tan* stands for field components tangent to the cylinder surface ($\hat{\boldsymbol{\varphi}}, \hat{\mathbf{z}}$). Assuming lossless, reciprocal and local surfaces renders $\underline{\underline{\mathbf{Y}}}_s, \underline{\underline{\mathbf{Z}}}_{ms}$ purely imaginary and symmetric and $\underline{\underline{\mathbf{a}}}, \underline{\underline{\mathbf{b}}}$ are real and satisfy $\underline{\underline{\mathbf{b}}} = -\underline{\underline{\mathbf{a}}}^T$ [32,33]. It should also be stressed that due to the complex surface responses that we study, in general the propagating waves do not possess a "pure" TE / TM nature, but rather a mixed state between the two.

### 3. Asymmetric propagation over folded hyperbolic metasurfaces

Surface wave propagation on hyperbolic metasurfaces have been shown to support many interesting phenomena, including broadband, highly confined surface waves, and enhanced local density of states. In the following, we show that the application of these properties in cylindrical tubes combines and couples them with orbital angular momentum, which can lead to valley selective propagation [34]. We pay particular attention to hyperbolic surfaces folded along various directions, analyzing their guiding properties as the folding angle $\theta$ is changed, which is incorporated into the equations by properly rotating the surface matrices $\underline{\underline{\mathbf{Y}}}_s, \underline{\underline{\mathbf{Z}}}_{ms}, \underline{\underline{\mathbf{a}}}, \underline{\underline{\mathbf{b}}}$ (see Appendix 1). We start by considering a planar surface with only a non-zero electric admittance tensor of the form

$$\underline{\underline{\mathbf{Y}}}_{s,p} = \begin{pmatrix} Y_{0,yy} & 0 \\ 0 & Y_{0,zz} \end{pmatrix} = \frac{j}{\eta_0} \begin{pmatrix} X_1 & 0 \\ 0 & X_2 \end{pmatrix}, \qquad (4)$$

Where throughout this work we will use matrices with "p" index to denote the response of the *planar* structures corresponding to the inset in figure (1) before the rotation, and without the "p" index for the resulting cylindrical structures. We then apply the rotation operators, and obtain the admittance tensor in the $\varphi - z$ coordinate system corresponding to the cylindrical geometry

$$\underline{\underline{\mathbf{Y}}}_s = \begin{pmatrix} Y_{\varphi\varphi} & Y_{\varphi z} \\ Y_{z\varphi} & Y_{zz} \end{pmatrix} = \frac{j}{\eta_0} \begin{pmatrix} X_1 \cos^2\theta + X_2 \sin^2\theta & (X_1 - X_2)\sin\theta\cos\theta \\ (X_1 - X_2)\sin\theta\cos\theta & X_1 \sin^2\theta + X_2 \cos^2\theta \end{pmatrix}. \tag{5}$$

It is important to note that $z$ direction before rotation and folding refers to one of the principle axes of the original planar surface, whereas after the folding it refers to the axial cylinder direction. We start by gradually extending the work in [20,21], by examining the special case of $X_1 = 0$. the dispersion equation becomes

$$\frac{I_n'(\tau a) K_n'(\tau a)}{I_n(\tau a) K_n(\tau a)} = -\frac{(n\beta a - a^2\tau^2 \cot\theta)^2}{k_0^2 a^4 \tau^2} - \frac{1 + \cot^2\theta}{k_0 a X_2 I_n(\tau a) K_n(\tau a)}, \tag{6}$$

Which collapses to the relation presented in [20,21] when $X_2 \to \pm\infty$. $I_n', K_n'$ are the first derivatives of the modified Bessel functions with respect to the argument. A dispersion plot is shown in figure (2a) for $\theta = \pi/4$, $X_2 = -4$ and since we have chosen an inductive surface with $X_2 < 0$, the modes are quasi-TM in nature. We see an interesting asymmetry in the band diagram associated with the OAM direction of the mode, where a specific mode *n* has different propagation features to opposite propagation directions. This asymmetry comes from the fact that the sheath-helix has a specific "handedness". When a mode *n* propagates in the opposite directions, the rotation direction of the phase flips as opposed to the handedness of the sheath helix which stays the same, making the waves and the helix interact differently. As expected based on time-reversal symmetry considerations, the $-n$ mode will have the opposite relations. For each higher-order mode, the tangency point of its dispersion with the $\beta a$ axis is shifted horizontally by approximately $\Delta \beta a = \tan \theta$, which can be seen by taking a $\beta \gg k_0$, low frequency limit of equation (6) in a similar manner to [21] (see additional remarks in Appendix 2 and Appendix 3) yielding $(\beta a \cot \theta - n)^2 = 0$. This case is a platform for interesting directional effects when such a waveguide interacts with a wave carrying angular momentum due to the asymmetry present.

This asymmetry can be pushed further if we examine a folded hyperbolic metasurface in the same configuration, considering $X_1 > 0$ (or in general the opposite sign of $X_2$) In Appendix 2 we give some of the guiding characteristics for a general relation between $X_1$ and $X_2$, where here, for simplicity, we will focus on the special case $|X_1| = |X_2| = X$ with $\theta = 45°$. The dispersion equation in this scenario becomes

$$\frac{1}{\tau a} - \tau a I_n K_n I_n' K_n' + \frac{2\beta n X}{\tau a} I_n K_n = 0,  \tag{7}$$

and the corresponding dispersion curves are shown in figure (2b) for $X_2 = -X = -4$. The asymmetry with respect to the mode index is more extreme here, and positive angular-momentum modes can propagate only to the right, whereas negative angular-momentum propagates to the left. This OAM-momentum coupling is reminiscent of the edge state spin-momentum coupling in photonic topological insulators [35] and in the vicinity of hyperbolic metamaterial slabs [36]. For $n = 0$ equation (7) has no real solutions for $\tau$, therefore the zero-index mode ($n = 0$) is not supported at all. Following this, if we try to excite this waveguide with OAM neutral excitation (any field / source pattern that would be invariant as a function of $\varphi$ in cylindrical coordinates), no propagation of waves will occur since the projection of such an excitation on the propagating modes will be zero [37]. This makes the waveguide completely dark to OAM neutral excitation. Each mode has a horizontal asymptote (red dashed lines in Fig. (2b)) in frequencies $f_{r,n}$ that satisfy (calculation remarks in Appendix 3)

$$\frac{2\pi f_{r,n} a}{c} = \frac{4|nX|}{4+X^2},\tag{8}$$

corresponding to surface waves highly confined to the tube. In this scenario, the modes have balanced TM and TE nature (fields of both polarizations are excited in comparable intensities), implying that they can strongly interact with a wide range of polarizations, with an ideal optical response to sort angular momentum and valley responses. Operating close to $f_{r,n}$, we expect highly mode-selective operation, due to the significant differences in mode confinement, and cutoff of the lower modes. To examine these opportunities, we consider two nearby localized emitters with same magnitude polarized along $\hat{\mathbf{z}}$, but with different phase, positioned in the location of the green vectors in figure (2c,2d). Such a source is used as a simple version of an excitation that couples efficiently to a specific value of orbital angular momentum $n$. Naturally, a mode $n$ would best couple to a current source of the form $J_z = J_{z0} e^{in\varphi}$, and our two-dipole source is a 2-point sampling of this optimal current distribution. The phase difference is chosen to couple to the $n=3$ mode, and the excitation frequency is $0.985 f_{r,3}$. We have used COMSOL electromagnetic solver to simulate the response of the studied structure to this excitation, where the emitters were modeled as electric point dipoles $\mathbf{p} = p\hat{\mathbf{z}}$. When the emitters excite the helical tube of figure (2a), figure (2c) shows that many modes are excited, as we do not see a clear wave pattern corresponding to a specific $n$. On the face of the cylinder we plot $E_z$, and on the planar panels perpendicular to the cylinder we plot the electric field intensity. The insets show the relative intensity of each mode $n$ [38], and no directionality is noticed. Figure (2d) shows the same excitation when applied to the folded hyperbolic metasurface, displaying highly mode-selective propagation and directionality. This can allow us to design emitters that strongly couple and launch specific angular momentum values using finite tube sections. Additional insights can be gained by examining the optical helicity density $\mathfrak{S}$ [39,40] for the modes supported by the folded hyperbolic metasurface, as calculated in Appendix 3. Since our interest is focused on the highly confined modes, this calculation can be greatly simplified by examining $\mathfrak{S}_\infty = \mathfrak{S}(\beta \gg k_0)$. In particular, it follows that $sign(\mathfrak{S}_\infty^{i,o}) = \pm sign(X)$. This can help customize the way our waveguide interacts with spin and helicity carrying sources such as valley excitons. In particular, one can envision directional sorting and routing of valley excitons using the aforementioned interactions when folded hyperbolic metasurfaces are coated with excitonic layers around the resonant frequencies of the various modes.

### 4. Asymmetry enhancement using magnetic responses

By virtue of duality, folded hyperbolic metasurfaces characterized by only magnetic impedance in (4), will present dual propagation features to the case analyzed in figure (2). Interesting opportunities arise if we combine these two responses, further enhancing the asymmetry and directionality of the supported modes. As a testbed, we first examine the case of a *planar* surface with the same electric response shown in equation (4), supplemented by

$$\underline{\underline{\mathbf{Z}}}_{ms,p} = \eta_0^2 \underline{\underline{\mathbf{Y}}}_{s,p} \;,\; \mathbf{a} = \mathbf{b} = \begin{pmatrix} 0 & \Omega \\ -\Omega & 0 \end{pmatrix}. \tag{9}$$

If $X_1, X_2$ are with opposite signs, and we combine it with a properly tailored value of $\Omega$, the modal asymmetry manifests through an interestingly one-sided field distribution of the propagating fields, as shown in figure (3a). The $\Omega$ response is required to tailor the cross-interaction between the electric and magnetic fields, which gives rise to this field asymmetry, as balanced electric and magnetic currents can excite highly asymmetric field distribution [41]. Bearing this example in mind, we notice (Appendix 1) that high order guided cylindrical modes contain an inherent and rich coupling between electric and magnetic fields, allowing us to achieve similar asymmetries in the cylindrical case, without the need for an $\Omega$-type response. All the cases we examine, are based on an electric-magnetic surface described (similarly to equation (4)) by

$$\underline{\underline{Y}}_{s,p} = \frac{j}{\eta_0}\begin{pmatrix} X & 0 \\ 0 & -X \end{pmatrix}, \quad \underline{\underline{Z}}_{ms,p} = j\eta_0\begin{pmatrix} X_m & 0 \\ 0 & -X_m \end{pmatrix}, \quad \mathbf{a} = 0, \quad \mathbf{b} = 0. \tag{10}$$

We start with the case of $X_m = X$, folded into a cylinder, with $\theta = 45°$. We find two types of solutions for each mode index $n$, corresponding to two distinct dispersion branches. The dispersion equations are

$$8nX\beta I_n K_n + \tau a\left[(X \mp 2)^2 I_n' K_n - (X \pm 2)^2 I_n K_n'\right] = 0, \tag{11}$$

where the upper and lower signs correspond to two distinct solution families. The dispersion curves are shown in figure (3b) for $X = 4$, and the modal fields satisfy

$$E_z^{i,o}(r=a) = \pm j\eta_0 H_z^{i,o}(r=a), \tag{12}$$

regardless of $X$, where the plus (minus) sign corresponds to the two distinct modal solutions. When substituting the dispersion solutions back into the governing equations (equations (23) in Appendix 1) to obtain the mode profile, the difference between the two modal solutions becomes apparent. The field amplitudes on the tube surface (both electric and magnetic, according to equation (12)) satisfy

$$\frac{|E_z^i|}{|E_z^o|} = \frac{|H_z^i|}{|H_z^o|} = \frac{(X \pm 2)}{(X \mp 2)} \qquad (13)$$

Yielding that the first set of modes is concentrated mostly in the outer side, while the second is on the inner region, corresponding to the labels in Figure (3b). This figure also shows that, by tuning the operation frequency to be slightly above or below $f_{r,n}$, we can control which field distribution is dominant. When $X = 2$, we maximize the contrast between inside and outside fields, yielding slow-wave modes concentrated purely in the interior or exterior part of the cylinder. In this extreme scenario, the modes not only spatially separated, but also totally spectrally separated – they are guided in completely different frequency bands, as shown in figure (3c). New cutoff frequencies at $k_{0,c} a = \sqrt{n(n+1)}$ are induced in this regime (blue lines in the figure). In this scenario, Eq. (12) ensures that $sign(\mathfrak{S}_\infty^{i,o}) = \mp 1$, coupling the helicity to the geometrical domain where the fields are concentrated.

Another interesting class corresponds to the same relation as in equation (10), this time with $X_m = -4/X$. As before we perform the folding along the $\theta = 45°$ line. The dispersion equation obtained in this case is completely symmetric for propagation to the left and right directions

$$|n\beta| = \frac{(k_0 a)(\tau a)}{4|X|} \sqrt{\left(4\frac{K'_n}{K_n} - \frac{I'_n}{I_n} X^2\right)\left(\frac{K'_n}{K_n} X^2 - 4\frac{I'_n}{I_n}\right)}, \qquad (14)$$

and the corresponding curves obtained are shown in Figure (4) for $X = 4$. The asymmetry in this case is revealed when closely examining the helicity of the propagating modes, satisfying $sign(\mathfrak{S}_\infty^{i,o}) = -sign(n\beta)$ and exhibiting strong helicity-momentum-OAM coupling. Additionally, it is worth mentioning the case of $X_m = -X$. Here, no such pronounced asymmetry is present, however it allows tailoring of the TE/TM ratio of the propagating waves. If we denote this ratio $\zeta$ (same notation is also used in Appendix 3) we obtain

$$|\zeta| = \frac{|E_z|}{|\eta_0 H_z|} = \left|\frac{X}{2}\right| \tag{15}$$

with dispersion curve very similar to the previous case, presented in figure (4).

## 5. Conclusions

In this paper, we have explored surface wave propagation over metasurface tubes. We have shown that folded hyperbolic metasurfaces can form an interesting platform for nanophotonics and valleytronics applications. They yield highly asymmetric propagation properties in terms of angular momentum, enabling largely unusual responses when properly tuning their impedance parameters. Around the resonance frequencies, the structures obtained may strongly interact with OAM / spin carrying waves such as the vortex Laguerre-Gauss beams [42] (with many more examples in [43]) or spin-specific excitons [44–46] which utilize the valley degree of freedom in transition metal dichalcogenides. When incorporating magnetic response of similar nature, These tubes also give rise to tunable asymmetry such as high field contrast between the inner and outer domains in addition to the asymmetric propagation. For all of the cases studied, strong OAM-momentum coupling was demonstrated which can be viewed as an extension of the spin-momentum coupling present in topological edge states. For each studied case, it was shown that the sign of the optical helicity for the propagating waves is controlled and coupled to various parameters – surface impedance, field propagation domain, OAM and wavenumber.

**Appendix 1**

To obtain the dispersion equation, we start by expressing the tangent fields using the coefficients $A^i, A^o, B^i, B^o$ defined in equation (1), arranged in a vector $\mathbf{C}^{i,o} = \left( A^{i,o}, B^{i,o} \right)^T$

$$\mathbf{E}^{i,o} = e^{jn\varphi} e^{-j\beta z} \underline{\underline{\mathbf{M}}}_{E,(i,o)} \mathbf{C}^{i,o} \ , \ \mathbf{H}^{i,o} = e^{jn\varphi} e^{-j\beta z} \underline{\underline{\mathbf{M}}}_{H,(i,o)} \mathbf{C}^{i,o} \tag{16}$$

and the matrices $\underline{\underline{\mathbf{M}}}_E, \underline{\underline{\mathbf{M}}}_H$ are defined as [21,31]

$$\mathbf{E}^i_{n,\tan}(r=a) = \begin{bmatrix} E^i_\varphi \\ E^i_z \end{bmatrix}_{r=a} = e^{jn\varphi - j\beta z} \begin{bmatrix} -\dfrac{n\beta}{\tau^2} I_n(\tau a) & -\dfrac{jk_0 a}{\tau} I_n{'}(\tau a) \\ I_n(\tau a) & 0 \end{bmatrix} \begin{bmatrix} A^i_n \\ B^i_n \end{bmatrix} = e^{jn\varphi - j\beta z} \underline{\underline{\mathbf{M}}}_{E,i} \mathbf{C}^i, \tag{17}$$

$$\mathbf{E}^o_{n,\tan}(r=a) = \begin{bmatrix} E^o_\varphi \\ E^o_z \end{bmatrix}_{r=a} = e^{jn\varphi - j\beta z} \begin{bmatrix} -\dfrac{n\beta}{\tau^2} K_n(\tau a) & -\dfrac{jk_0 a}{\tau} K_n{'}(\tau a) \\ K_n(\tau a) & 0 \end{bmatrix} \begin{bmatrix} A^o_n \\ B^o_n \end{bmatrix} = e^{jn\varphi - j\beta z} \underline{\underline{\mathbf{M}}}_{E,o} \mathbf{C}^o, \tag{18}$$

$$\mathbf{H}^i_{n,\tan}(r=a) = \begin{bmatrix} H^i_\varphi \\ H^i_z \end{bmatrix}_{r=a} = e^{jn\varphi - j\beta z} \begin{bmatrix} \dfrac{jk_0 a}{\tau} I_n{'}(\tau a) & -\dfrac{n\beta}{\tau^2} I_n(\tau a) \\ 0 & I_n(\tau a) \end{bmatrix} \begin{bmatrix} A^i_n \\ B^i_n \end{bmatrix} = e^{jn\varphi - j\beta z} \underline{\underline{\mathbf{M}}}_{H,i} \mathbf{C}^i, \tag{19}$$

$$\mathbf{H}^o_{n,\tan}(r=a) = \begin{bmatrix} H^o_\varphi \\ H^o_z \end{bmatrix}_{r=a} = e^{jn\varphi - j\beta z} \begin{bmatrix} \dfrac{jk_0 a}{\tau} K_n{'}(\tau a) & -\dfrac{n\beta}{\tau^2} K_n(\tau a) \\ 0 & K_n(\tau a) \end{bmatrix} \begin{bmatrix} A^o_n \\ B^o_n \end{bmatrix} = e^{jn\varphi - j\beta z} \underline{\underline{\mathbf{M}}}_{H,o} \mathbf{C}^o. \tag{20}$$

To easily substitute these relations into the boundary conditions, we represent the $\hat{\mathbf{r}} \times$ operation that would be needed in equation (2) and (3) using a $2 \times 2$ matrix $\hat{\mathbf{r}} \times \mathbf{V}_{\tan} = \underline{\underline{\mathbf{N}}}_r \mathbf{V}_{\tan}$ (where $\mathbf{V}_{\tan}$ is any tangential column vector of the form $\mathbf{V}_{\tan} = [V_\varphi, V_z]^T$, and $\underline{\underline{\mathbf{N}}}_r = \begin{pmatrix} 0 & -1 \\ 1 & 0 \end{pmatrix}$). We also define the unitless response matrices $\underline{\underline{\mathbf{X}}}_s, \underline{\underline{\mathbf{X}}}_{ms}$

$$\underline{\underline{\mathbf{Y}}}_s = j\underline{\underline{\mathbf{X}}}_s / \eta_0 \ , \ \underline{\underline{\mathbf{Z}}}_{ms} = j\underline{\underline{\mathbf{X}}}_{ms} \eta_0. \tag{21}$$

To express the boundary conditions, we rotate each of the fields by the angle $\theta$ with respect to $\hat{\mathbf{z}}$, and then rotate the result back to the lab frame. This results in rotated response matrices of the form $\underline{\underline{\mathbf{m}}}^\theta = \underline{\underline{\mathbf{R}}}^\theta \underline{\underline{\mathbf{m}}} \underline{\underline{\mathbf{R}}}^{-\theta}$, where $\underline{\underline{\mathbf{m}}}$ is any of tensors defined in the constitutive relations in equations (2) and (3), and $\underline{\underline{\mathbf{R}}}^\theta$ is the 2D rotation matrix by angle $\theta$. By substituting these into the boundary conditions in equations (2) and (3) we obtain

$$\underline{\underline{\mathbf{M}}}\mathbf{C} = \begin{bmatrix} \tfrac{1}{2}\underline{\underline{\mathbf{X}}}_s^\theta \underline{\underline{\mathbf{M}}}_{E,i} + \left(\tfrac{1}{2}\underline{\underline{\mathbf{a}}}^\theta + \underline{\underline{\mathbf{N}}}_r\right)\underline{\underline{\mathbf{M}}}_{H,i} & \tfrac{1}{2}\underline{\underline{\mathbf{X}}}_s^\theta \underline{\underline{\mathbf{M}}}_{E,o} + \left(\tfrac{1}{2}\underline{\underline{\mathbf{a}}}^\theta - \underline{\underline{\mathbf{N}}}_r\right)\underline{\underline{\mathbf{M}}}_{H,o} \\ \left(\tfrac{1}{2}\underline{\underline{\mathbf{b}}}^\theta - \underline{\underline{\mathbf{N}}}_r\right)\underline{\underline{\mathbf{M}}}_{E,i} + \tfrac{1}{2}\underline{\underline{\mathbf{X}}}_{ms}^\theta \underline{\underline{\mathbf{M}}}_{H,i} & \left(\tfrac{1}{2}\underline{\underline{\mathbf{b}}}^\theta + \underline{\underline{\mathbf{N}}}_r\right)\underline{\underline{\mathbf{M}}}_{E,o} + \tfrac{1}{2}\underline{\underline{\mathbf{X}}}_{ms}^\theta \underline{\underline{\mathbf{M}}}_{H,o} \end{bmatrix} \begin{bmatrix} \mathbf{C}^i \\ \mathbf{C}^o \end{bmatrix} = 0. \tag{22}$$

For a non-trivial solution to exist, we need $Det\{\underline{\underline{\mathbf{M}}}\} = 0$, providing the dispersion equation for propagating waves.

**Appendix 2**

To examine a more general folded hyperbolic metasurface, we start by defining $X_2 = X, X_1 = -sX$ in equation (4). The critical fold angle $\theta_c$ which gives the same type of dispersion curved shown for the simpler cases corresponds to the condition that the $zz$ component of the rotated admittance tensor is zero (only when both $X_1, X_2 \neq 0$). This angle satisfies

$$\theta_c = \text{arccot}\left[\sqrt{s}\right]. \tag{23}$$

After some long, but straightforward algebra we obtain the dispersion equation

$$2nX\beta a\sqrt{s}I_n K_n + k_0 a\left(1 - sX^2\tau^2 a^2 I_n' I_n K_n K_n'\right) + \frac{X(1-s)}{\tau^2 a^2}\left(n^2\beta^2 a^2 I_n K_n + k_0^2 a^4 \tau^2 I_n' K_n'\right) = 0 \tag{24}$$

And if we substitute $s = 1$ we will obtain equation (7). The series of resonance frequencies for each mode index $n$ can be written in a similar way to equation (8)

$$\frac{2\pi f_{r,n} a}{c} = \frac{4\sqrt{s}|nX|}{4 + sX^2} \tag{25}$$

And a typical form of the dispersion is shown in figure (5) for two cases. The expressions for the resonance frequencies are obtained using the large argument approximated forms for the modified Bessel functions in the dispersion equations [47]. In both cases presented in figure (5), the values were chosen in a way that the off-diagonal terms in the rotated admittance tensor defined in equation (4) of the main text will be the same. The difference, therefore, stems from the fact that in the general case the other diagonal term ($\varphi\varphi$) is not zero, and their value and sign determine the behavior of the additional dispersion branches.

## Appendix 3

We use the helicity density $\mathfrak{S}$ definition from [39,40], which reads for our case

$$\mathfrak{S} = \frac{1}{2\omega} \text{Im}\{\mathbf{H}^* \cdot \mathbf{E}\} \tag{26}$$

In addition, for all cases examined in the main text, the relation between the TE and TM components can be characterized by

$$\eta_0 B_n^{i,o} = j\zeta_{i,o} A_n^{i,o} \tag{27}$$

with $\zeta \in \mathbb{R}$ being a unitless proportionality constant, and $B_n^{i,o}, A_n^{i,o}$ defined in equation (1). This lets us write the helicity density of the guided fields as

$$\mathfrak{S}^i = -\frac{|A_n^i|^2}{2\omega \bar{\tau}^4} \left[ \zeta_i I_n^2(\bar{\tau}) \left[ \bar{\tau}^4 + n^2 \left( \bar{\beta}^2 + \bar{a}^2 \right) \right] + \zeta_i I_n'^2(\bar{\tau}) \bar{\tau}^2 \left( \bar{\beta}^2 + \bar{a}^2 \right) - 2 I_n I_n' \, n \bar{\beta} \bar{\tau} \bar{a} \left( 1 + \zeta_i^2 \right) \right]$$

$$\mathfrak{S}^o = -\frac{|A_n^o|^2}{2\omega \bar{\tau}^4} \left[ \zeta_i K_n^2(\bar{\tau}) \left[ \bar{\tau}^4 + n^2 \left( \bar{\beta}^2 + \bar{a}^2 \right) \right] + \zeta_i K_n'^2(\bar{\tau}) \bar{\tau}^2 \left( \bar{\beta}^2 + \bar{a}^2 \right) - 2 K_n K_n' \, n \bar{\beta} \bar{\tau} \bar{a} \left( 1 + \zeta_i^2 \right) \right]$$

$$(28)$$

with $\bar{\beta} = \beta a$, $\bar{\tau} = \tau a$, $\bar{a} = k_0 a$. In our work we study the highly confined waves, with large wave numbers, letting us simplify this expression significantly

$$\mathfrak{S}_\infty^i = -\frac{|A_n^i|^2 \zeta_i}{2\omega} \left[ I_n^2(\bar{\tau}) + I_n'^2(\bar{\tau}) \right],$$

$$\mathfrak{S}_\infty^o = -\frac{|A_n^i|^2 \zeta_i}{2\omega} \left[ K_n^2(\bar{\tau}) + K_n'^2(\bar{\tau}) \right]$$

$$(29)$$

based on large-argument approximations of the modified Bessel functions [47], where we used the symbol $\mathfrak{S}_\infty = \mathfrak{S}(\beta \gg k_0)$. This result is valid as long as $\zeta$ is not strongly dependent on $\bar{\beta}$, which is indeed the case in all of the systems examined. It is important to note that for any value of $\bar{\tau}$ the expressions in the square brackets are positive, therefore, the sign of the helicity density is determined solely by the sign of $\zeta$, and satisfies

$$sign\left(\mathfrak{S}_{\infty}^{i,o}\right) = -sign\left(\zeta^{i,o}\right) \qquad (30)$$

**Figures**

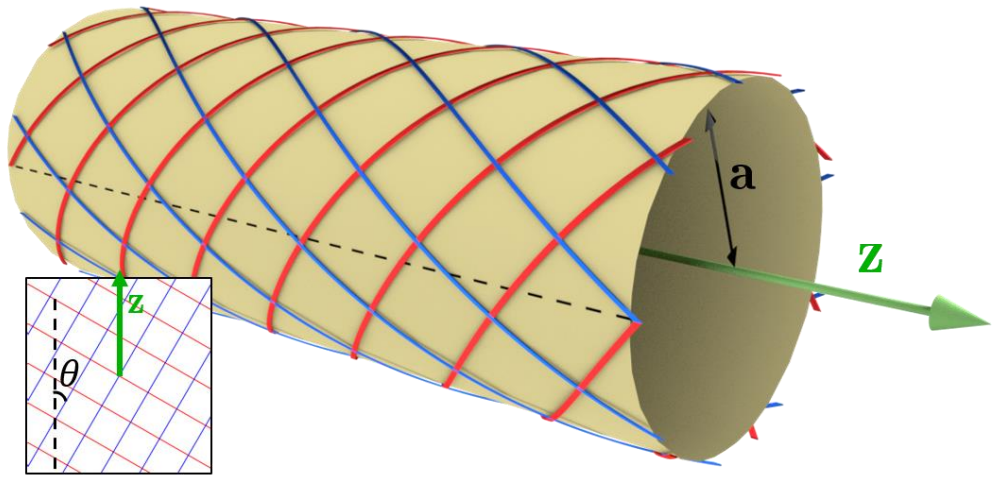

**Figure 1**. The geometry under analysis, showing the principal axes of the constitutive relations tensors. When rotated, they form a skewed reference system over the surface of the cylinder.

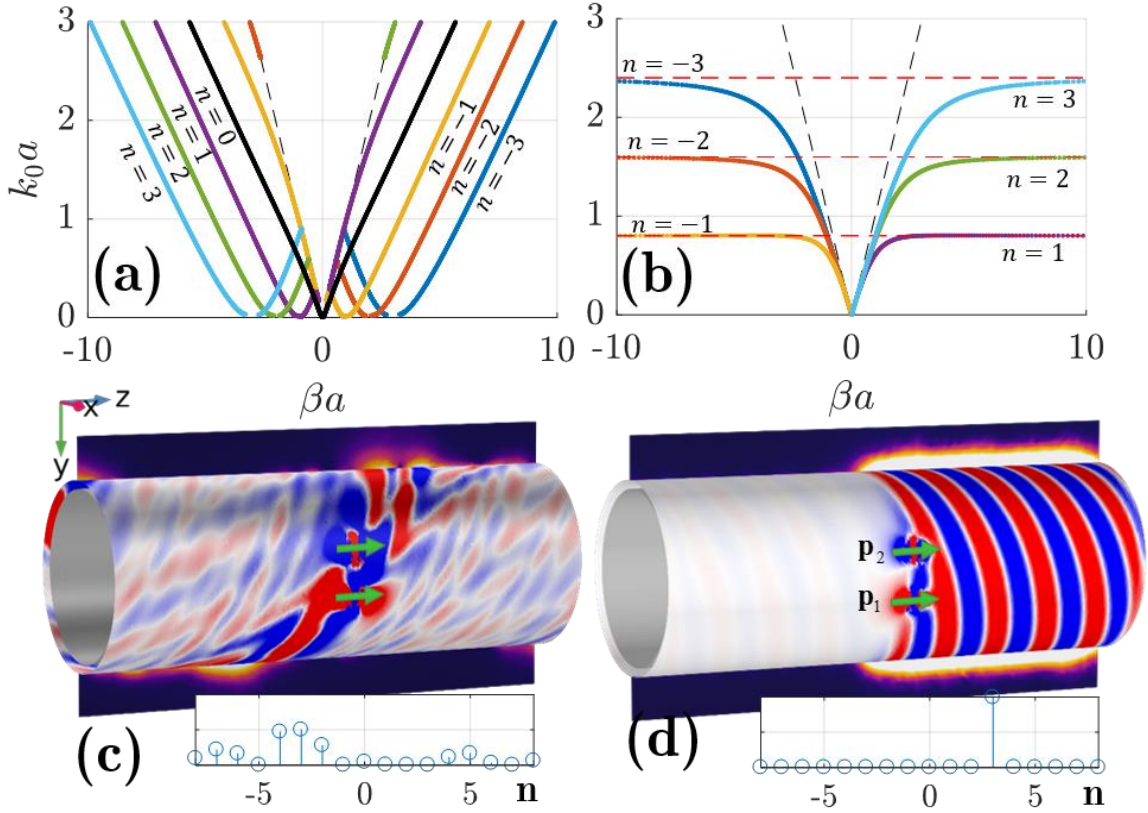

**Figure 2**. (a) Surface wave dispersion curves when only $\sigma_{zz}$ in equation (4) is present with $X_2 = -4$ and $\theta = \pi/4$. Dashed black lines show the light-lines, $k_0 = \omega/c$. (b) Surface wave dispersion curves for a "folded" hyperbolic metasurface with $X_1 = 4, X_2 = -4$, Dashed black lines shown the light-lines. The mode $n = 0$ is not present, as it does not propagate. (c) On the cylindrical envelope we show $E_z$ as sampled close to the impedance cylinder, when exciting with two dephased electric dipoles. The cylinder properties correspond to the dispersion in panel (a). The black, perpendicular panels show the electric field intensity, where we see the fast decay as the distance from the cylinder increases. The inset shows the modal content of the fields near the cylinder, where we see a mixture of many modes. (d) same as (c), but corresponding to the cylinder

properties matching the dispersion in panel (b). highly directional excitation is observed, and pronounced mode selectivity with a single mode propagating ($n = 3$).

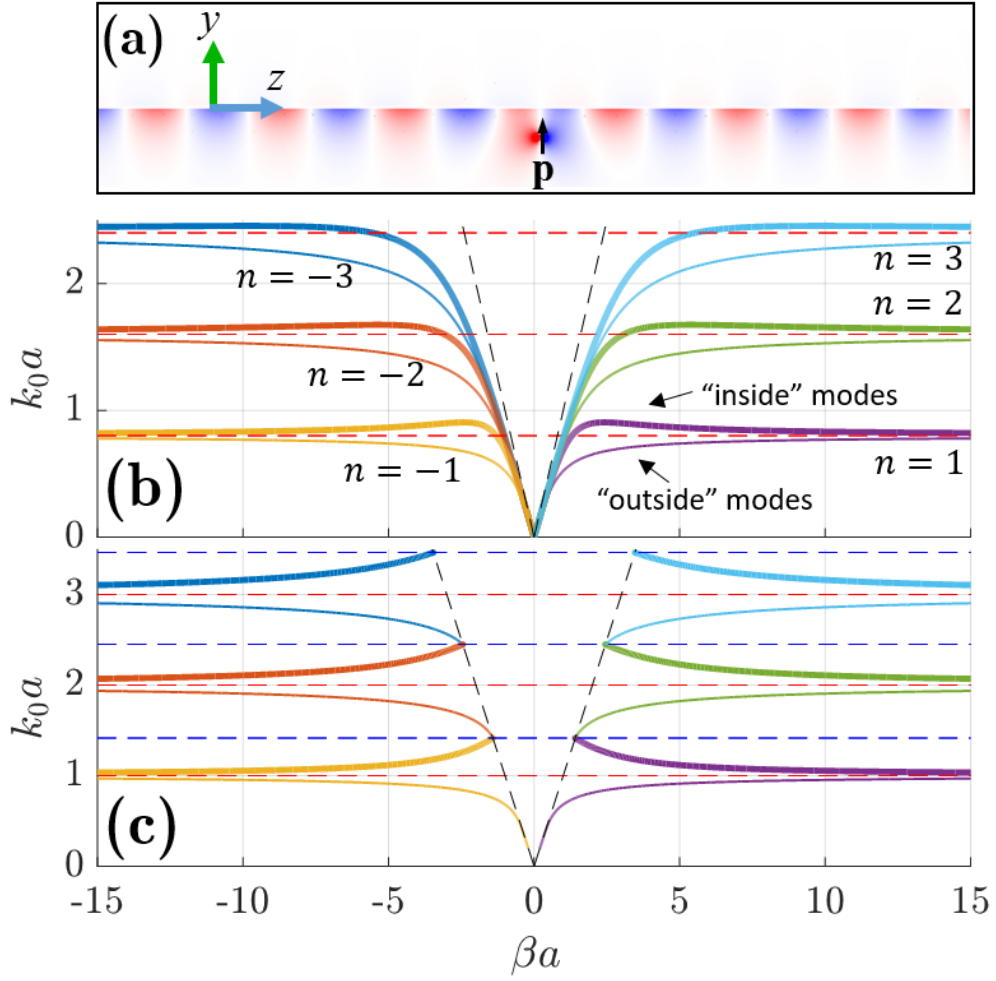

**Figure 3**. (a) Excitation of surface waves over a *planar* surface with $Y_{0,yy} = -j0.3/\eta_0$, $Y_{0,zz} = j16.133/\eta_0$ and $\Omega = 0.2$, corresponding to equation (9). The modal fields are concentrated at the bottom surface. (b) Surface wave dispersion curves for $X = X_m = 4$ in equation (10). Dashed black lines are the light lines and $k_0 = \omega/c$ (c) Same as (b) but with $X = X_m = 2$.

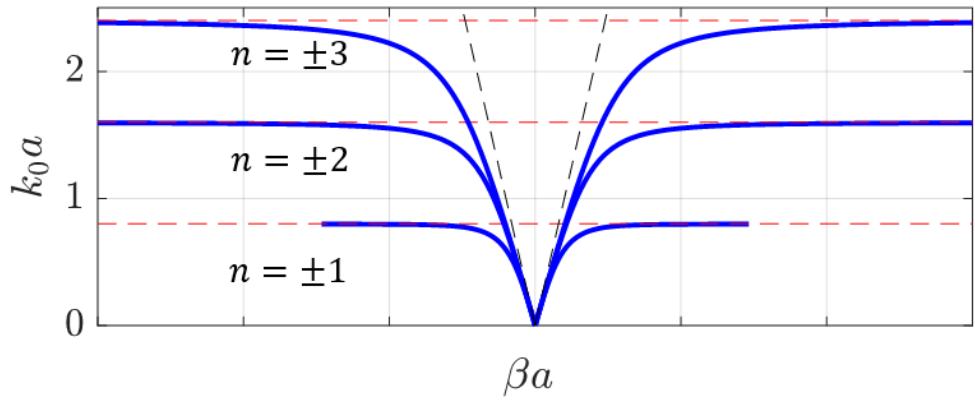

**Figure 4.** Dispersion curves for the case of $X_m = -4/X$ in equation (10). The parameters are $X = 4, X_m = -1$. Black dashed lines are the light lines, and $k_0 = \omega/c$.

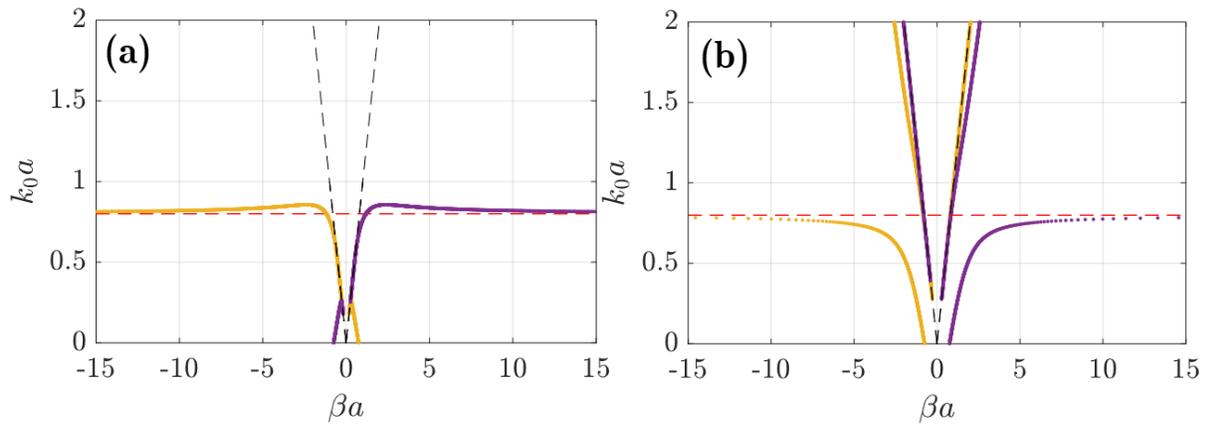

**Figure 5**. (a) Dispersion for a general folded hyperbolic metasurface for $\theta = \theta_c$ from equation (23), with $s = 0.25$, $X = -2$ (appendix 2). Red dashed line shows the value of $k_0 a$ corresponding to $f_{c,1}$. (b) the same as a, this time with $s = 4, X = -0.5$.